\documentclass[%
  reprint,
  superscriptaddress,
  nofootinbib,
  amsmath,amssymb,
  aps,
  prl,
  floatfix,
]{revtex4-2}

\usepackage{graphicx}
\usepackage{dcolumn}
\usepackage{bm}
\usepackage{comment}
\usepackage{amsmath}
\usepackage{amsfonts}
\usepackage{amssymb}
\usepackage{mathtools}
\usepackage[utf8]{inputenc}
\usepackage{nicefrac}
\usepackage[english]{babel}
\usepackage{siunitx}
\usepackage{orcidlink}
\usepackage{multirow}
\usepackage{booktabs}
\usepackage[normalem]{ulem}

\usepackage[svgnames]{xcolor}
\definecolor{linkColor}{rgb}{0.25,0.55,0.85}
\usepackage{hyperref}
\hypersetup{
  colorlinks  = true,
  allcolors   = linkColor,
  pdfborder   = {0 0 0},
  pdfencoding = auto
}
\usepackage[capitalise]{cleveref}

\renewcommand{\cite}{\citep}

\newcommand{\AMBm}{\mathrm{AMB}^{-}}
\newcommand{\AMBp}{\mathrm{AMB}^{+}}

\begin{document}

\title{Active Model B$^{-}$ from Mass-Conserving  Reaction-Diffusion Systems}

\author{Davide Toffenetti\,\orcidlink{0009-0001-3698-7586}}
\thanks{These authors contributed equally to this work.}
\affiliation{Arnold Sommerfeld Center for Theoretical Physics and Center for NanoScience,
Department of Physics, Ludwig-Maximilians-Universit\"at M\"unchen,
Theresienstra\ss e~37, D-80333 M\"unchen, Germany}

\author{Beatrice Nettuno\,\orcidlink{0009-0000-8564-0004}}
\thanks{These authors contributed equally to this work.}
\affiliation{Arnold Sommerfeld Center for Theoretical Physics and Center for NanoScience,
Department of Physics, Ludwig-Maximilians-Universit\"at M\"unchen,
Theresienstra\ss e~37, D-80333 M\"unchen, Germany}

\author{Henrik Weyer\,\orcidlink{0000-0001-9612-0741}}
\affiliation{Arnold Sommerfeld Center for Theoretical Physics and Center for NanoScience,
Department of Physics, Ludwig-Maximilians-Universit\"at M\"unchen,
Theresienstra\ss e~37, D-80333 M\"unchen, Germany}
\affiliation{Kavli Institute for Theoretical Physics, University of California,
Santa Barbara, CA~93106, USA}

\author{Erwin Frey\,\orcidlink{0000-0001-8792-3358}}
\email[Contact author: ]{frey@lmu.de}
\affiliation{Arnold Sommerfeld Center for Theoretical Physics and Center for NanoScience,
Department of Physics, Ludwig-Maximilians-Universit\"at M\"unchen,
Theresienstra\ss e~37, D-80333 M\"unchen, Germany}
\affiliation{Max Planck School Matter to Life, Hofgartenstra\ss e~8,
D-80539 M\"unchen, Germany}

\date{\today}

\begin{abstract}
We show that the  late-time dynamics of a minimal three-component mass-conserving reaction--diffusion system reduce to a scalar active field theory, Active Model B$^-$ ($\AMBm$), in which a density-dependent interfacial coefficient $\kappa(\phi)$ turns negative at high density. This drives a finite-wavelength instability and stabilises microphase-separated patterns, in contrast to the unbounded coarsening of two-component mass-conserving systems. Unlike Active Model B$^+$, $\AMBm$ retains a chemical potential that remains a state function, inherited from the underlying conservation law, but admits no equation of state for the pressure.
\end{abstract}

\maketitle

Protein pattern formation under mass conservation is a fundamental mechanism of intracellular self-organisation~\cite{Frey_Weyer:2026, Goryachev_Leda:2017, EdelsteinKeshet_Dutot:2013}
One paradigmatic example is the Min system of \emph{E.~coli}, where MinD and MinE proteins undergo an ATP-driven membrane-binding cycle that positions the division machinery at mid-cell~\cite{Lutkenhaus:2007}.
In vitro, Min proteins form a remarkable diversity of self-organised patterns, including travelling waves and spirals~\cite{Loose_Schwille:2008}; mushrooms, amoebas, and bursts~\cite{Vecchiarelli_Mizuuchi:2016}; and quasi-stationary dots, stripes, and foam-like meshes~\cite{Glock_Schwille:2018}; with no progressive coarsening over experimental timescales.
Explaining wavelength and pattern selection within a strictly mass-conserving framework remains a central theoretical challenge.
Mass-conserving reaction--diffusion (McRD) systems provide the
minimal bottom-up description of pattern formation under a global
conservation law~\cite{Otsuji_Kuroda:2007, Goryachev_Pokhilko:2008,
Mori_EdelsteinKeshet:2008, Halatek_Frey:2018, Brauns_Frey:2020,
Frey_Weyer:2026}. Two-component systems generically coarsen
without bound~\cite{Ishihara_Mochizuki:2007, Tateno_Ishihara:2021,
Brauns_Frey:2021, Weyer_Frey:2023, Weyer_Leung_Frey:2026}, with
long-wavelength dynamics equivalent to Cahn--Hilliard or conserved
Allen--Cahn flows~\cite{Weyer_Frey:2023}, driven by mass redistribution
between domains independently of reaction kinetics~\cite{Brauns_Frey:2021}.
Finite-wavelength selection has been recovered along two routes within
McRD: weak breaking of mass conservation, where a unified mesoscopic
mechanism is known~\cite{Kolokolnikov_Wei:2006, Kolokolnikov_Wei:2007,
Brauns_Frey:2021, Weyer_Frey:2023}, and addition of a third reactive
species, where the phenomenology has been demonstrated case by
case~\cite{Jacobs_Deinum:2019, Chiou_Lew:2021, Wigbers_Fakhri:2021}.
A mechanism-level account of the second route, analogous to that of
the first, is missing.

A complementary top-down perspective on driven self-organisation is provided by scalar active field theories with a conserved order parameter~\cite{Cates_Nardini:2025}.
The minimal extension of Cahn--Hilliard dynamics by a nonequilibrium contribution to the chemical potential, Active Model~B~\cite{Wittkowski_Cates:2014}, generically coarsens to bulk phase separation.
Finite-wavelength selection can be recovered by coupling Cahn--Hilliard dynamics to chemical reactions that weakly break the conservation of the order parameter~\cite{Glotzer_Muthukumar:1995,Zwicker_Juelicher:2015} , or, while preserving conservation, by adding a non-gradient current as in Active Model~B$^+$~\cite{Tjhung_Cates:2018}, which reverses Ostwald ripening and stabilises microphase separation. Active Brownian particles establish one microscopic route to such field theories~\cite{Stenhammar_Cates:2013}, and their nonequilibrium structure is well understood~\cite{Nardini_Cates:2017,Caballero_Cates:2018,Cates_Nardini:2025}.
Whether mass-conserving biochemical reaction networks, the natural setting for intracellular pattern formation, give rise to active-field-theory dynamics is largely unexplored.

Both questions admit a common answer.
We consider a minimal three-component mass-conserving reaction-diffusion model in which an inactive cytosolic reservoir is coupled to the active dynamics by a reactivation rate $\nu$, motivated by nucleotide exchange in ATPase cycles.
Adiabatic elimination of the fast variables yields a closed equation for the conserved total density~$\phi$.
In the fast-reactivation limit ($\nu\to\infty$) this recovers passive Cahn--Hilliard dynamics; in the slow reactivation limit ($\nu\to 0$) it maps to an activator–substrate model \cite{Murray:2003,Gierer_Meinhardt:1972}, while at finite $\nu$ it gives a distinct active scalar field theory, Active Model~B$^-$ ($\AMBm$, Eq.~\eqref{eq:AMBminus}), a minimal active field theory which supports finite-length instability. In this theory, activity enters through a density-dependent interfacial coefficient $\kappa(\phi)$ that becomes negative at high density and explains the finite-wavelength selection observed in the three-component model.
Crucially, this is achieved within a purely gradient-current structure: unlike $\AMBp$, which recovers wavelength selection by introducing a non-gradient active current, $\AMBm$ preserves the gradient form of the current and locates activity entirely in $\kappa(\phi)$.
The reactivation rate $\nu$ interpolates continuously between regimes where patterns coarsen akin to the Cahn--Hilliard model, and a regime of stable
micro-phase-separated patterns (dots, stripes, foams) in agreement with Min system phenomenology.

\begin{figure*}[htb]
  \centering
  \includegraphics[width=\textwidth]{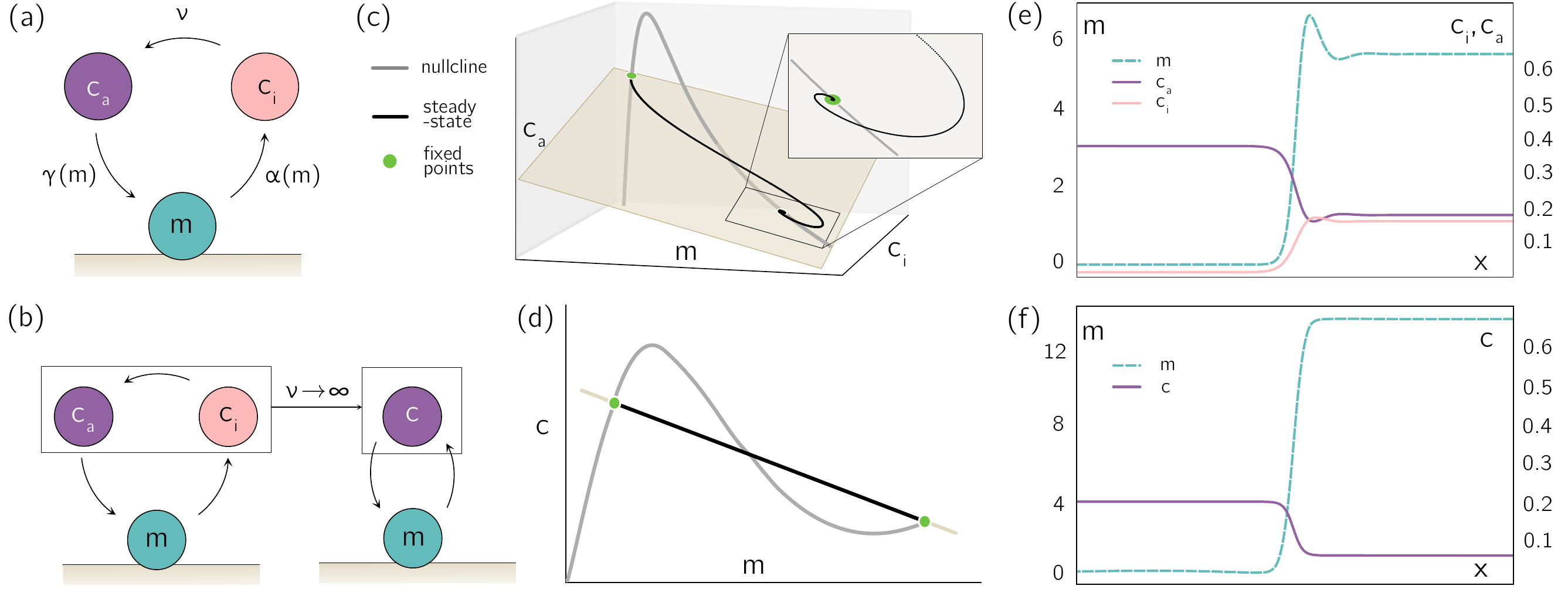}
\caption{%
(a) Reaction scheme of the three-component model
[Eq.~\eqref{eq:rde-three-comp}]: the membrane-bound species $m$ detaches
at rate $\alpha(m)$ into the inactive cytosolic state $c_{\mathrm{i}}$,
which reactivates at rate $\nu$ into the active state $c_{\mathrm{a}}$,
and $c_{\mathrm{a}}$ recruits back to the membrane at rate $\gamma(m)$.
(b) Limiting two-component mass-conserving model,
obtained in the limit ${\nu\to\infty}$, in which $c_{\mathrm{i}}$ and
$c_{\mathrm{a}}$ collapse onto a single lumped cytosolic species $c$.
(c) Phase-space construction for the three-component model: the
reactive nullcline (gray curve) defines the local
reactive equilibria, and the steady-state interface trajectory (black)
is constrained to a \emph{plane} of constant $\eta$ (beige, light
gray); the inset zooms in on the homogeneous fixed points
(green dots, gray dots).
(d) Phase-space construction for the two-component model: the
steady-state trajectory (black) is confined to a \emph{line} of
constant $\eta$ (beige, light gray); the
homogeneous fixed points are marked as dots (green, gray).
(e) Non-monotonic stationary interface profile of the three-component
model, corresponding to the trajectory in (c); $m$ (dashed teal) on left axis, $c_{\mathrm{a}}$ (solid purple, dark gray) and $c_{\mathrm{i}}$
(solid pink, light gray) on right axis.
(f) Monotonic interface profile corresponding to the trajectory in (d),
characteristic of two-component mass-conserving systems; $m$ (dashed)
and lumped $c$ (solid) shown.}
 \label{fig:model}
\end{figure*}

\paragraph{Three-component model and its limiting cases.}
We consider a minimal three-component mass-conserving reaction--diffusion model [Fig.~\ref{fig:model}(a)] with one membrane-bound species~$m$ and two cytosolic species:
an active component~$c_{\mathrm{a}}$ and an inactive component~$c_{\mathrm{i}}$  \cite{Chiou_Lew:2021}.
The three species form a unidirectional cycle $c_{\mathrm{a}} \to m \to c_{\mathrm{i}} \to c_{\mathrm{a}}$:
$c_{\mathrm{a}}$ attaches to the membrane at rate~$\gamma(m)$, membrane-bound proteins detach into the inactive state with a turnover flux~$\alpha(m)$, and the inactive species reactivates at rate~$\nu$.
The equations of motion read
\begin{subequations}
\label{eq:rde-three-comp}
\begin{align}
  \partial_t c_{\mathrm{a}}
  &= D_{\mathrm{c}}\,\nabla^2 c_{\mathrm{a}}
     + \nu\,c_{\mathrm{i}}
     - \gamma(m)\,c_{\mathrm{a}}\,,
  \label{eq:active-cytosol}\\
  \partial_t c_{\mathrm{i}}
  &= D_{\mathrm{c}}\,\nabla^2 c_{\mathrm{i}}
     - \nu\,c_{\mathrm{i}}
     + \alpha(m)\,,
  \label{eq:inactive-cytosol}\\
  \partial_t m
  &= D_{\mathrm{m}}\,\nabla^2 m
     + \gamma(m)\,c_{\mathrm{a}}
     - \alpha(m)\,,
  \label{eq:membrane-bound}
\end{align}
\end{subequations}
with cytosolic and membrane diffusivities $D_{\mathrm{c}} \gg D_{\mathrm{m}}$.
The total density ${\phi = c_{\mathrm{a}} + c_{\mathrm{i}} + m}$ obeys a continuity equation:
${\partial_t\phi = D_{\mathrm{c}}\,\nabla^2\eta}$, where
\begin{equation}
  \eta \coloneqq c_{\mathrm{a}} + c_{\mathrm{i}} + d_{\mathrm{m}}\,m\,,
  \qquad
  d_{\mathrm{m}} = D_{\mathrm{m}}/D_{\mathrm{c}}\,,
  \label{eq:total-mass-conservation}
\end{equation}
is the mass-redistribution potential~\cite{Ishihara_Mochizuki:2007, Halatek_Frey:2018, Brauns_Frey:2020}.

The rate~$\nu$ sets a diffusion length ${ \sqrt{D_{\mathrm{c}}/\nu}}$ for the inactive species and interpolates between two well-studied limits.
For ${\nu \to \infty}$, $c_{\mathrm{i}}$ vanishes and the model reduces to a two-component mass-conserving system in $c_{\mathrm{a}} + m \approx \phi$~\cite{Brauns_Frey:2020} [Fig.~\ref{fig:model}(b)].
For ${\nu \to 0}$, the inactive species acts as a homogeneous reservoir feeding active proteins into the cytosol at a constant rate, breaking mass conservation and recovering an  activator--substrate model~\cite{Meinhardt:2008,Brauns_Frey:2021,Weyer_Frey:2023} (see Supplemental Material~\cite{SM}).
At steady state, ${\partial_t\phi = 0}$ implies that the mass-redistribution potential is spatially uniform, ${\eta = \eta_\mathrm{stat}}$.
The geometric implication of this single constraint differs sharply between the two- and three-component cases.
In the two-component limit, the trajectory is confined to a \emph{line} in the $(m,c_{\mathrm{a}})$ plane [Fig.~\ref{fig:model}(d)], fixing all concentrations as functions of $\phi$.
The coexistence densities $\phi_\pm$ then follow from two profile-independent conditions, equal $\eta$ and turnover balance~\cite{Brauns_Frey:2020}, and the interface is necessarily monotonic [Fig.~\ref{fig:model}(f)].
In the three-component case, the same constraint $\eta = \eta_\mathrm{stat}$ defines a \emph{plane} in $(m,c_{\mathrm{a}},c_{\mathrm{i}})$ space [Fig.~\ref{fig:model}(c)], and the trajectory gains an internal degree of freedom.
The inactive species~$c_{\mathrm{i}}$ is no longer fixed locally by~$\phi$, but by the screened-diffusion balance,
\begin{equation}
  0 = D_{\mathrm{c}}\,\nabla^2 c_{\mathrm{i}} - \nu \, c_{\mathrm{i}} + \alpha(m)\,,
  \label{eq:ci-steady}
\end{equation}
permitting non-monotonic interface profiles [Fig.~\ref{fig:model}(e)].
The turnover-balance condition then becomes profile-dependent (End Matter) and ceases to be a coexistence condition that uniquely fixes the binodal values $\phi_{\pm}$, thereby allowing for a wider range of patterns.

\paragraph{Reduction to a scalar field theory.}
The conserved density $\phi$ evolves diffusively, while $\eta$ and $c_{\mathrm{i}}$ relax rapidly through reactions and can be eliminated adiabatically.
This requires the system to be in the late-time, diffusion-limited regime, where mass transport is slow compared with the local reactions~\cite{Weyer_Frey:2023}.
We expand the interface profile about the plane of constant $\eta$ rather than about the reactive nullcline as in Ref.~\cite{Robinson2025}.
A gradient expansion of the cytosolic field, valid for ${\nu \gg 1}$, supplemented by the closure ${\phi = m + \mathcal{O}(\nu^{-1}, d_{\mathrm{m}})}$, valid for ${d_{\mathrm{m}}\ll 1}$ (End Matter), yields a closed equation for $\phi$,
\begin{equation}
  \partial_t\phi
  =
  M\,\nabla^2
  \left[
    \eta^*
    - \kappa \,\nabla^2\phi
    + \lambda \,(\nabla\phi)^2
    + \nabla^4\chi
  \right] \, ,
  \label{eq:AMBminus}
\end{equation}
which we term Active Model B$^-$ ($\AMBm$).
The mobility $M=D_{\mathrm{c}}$ and the density-dependent coefficients $\eta^* (\phi)$, $\kappa (\phi)$, $\lambda (\phi)$, $\chi (\phi)$ are fixed by the microscopic rates (Table~\ref{tab:params}).
The superscript in $\AMBm$ reflects the sign change of $\kappa(\phi)$ at large densities, the defining feature of the model and the source of its finite-wavelength instability.
This complements $\AMBp$~\cite{Tjhung_Cates:2018}, which extends Cahn--Hilliard via a non-gradient current while keeping $\kappa$ positive.

\begin{table}[htb]
\centering
\setlength{\tabcolsep}{3pt}
\begin{tabular*}{\columnwidth}{@{\extracolsep{\fill}}lccccc}
\hline\hline
\noalign{\vskip 4pt}
$\AMBm$ & 3-comp.\ & 2-comp.\ &  AMB & PFC \\[4pt]
\hline
\noalign{\vskip 4pt}
$\eta^*(\phi)$
  & $\alpha/\gamma + \alpha /\nu + d_{\mathrm{m}}\phi$
  & $\alpha/\gamma + d_{\mathrm{m}}\phi$
  & $-\phi + \phi^3$
  & $-\phi + \phi^3$ \\[8pt]
$\kappa(\phi)$
  & $D_{\mathrm{m}}/\gamma - D_c\,\alpha'/\nu^2$
  & $D_{\mathrm{m}}/\gamma$
  & $\kappa>0$
  & $\kappa<0$ \\[8pt]
$\chi(\phi)$
  & $D^2_c\,\alpha/\nu^2$
  & $0$
  & $0$
  & $K\phi,\ K>0$ \\[8pt]
$\lambda(\phi)$
  & $D_c\,\alpha''/\nu^2$
  & $0$
  & $\lambda$
  & $0$ \\[8pt]
\hline\hline
\end{tabular*}
\caption{%
Coefficients of $\AMBm$ [Eq.~\eqref{eq:AMBminus}] in the four cases compared. The 3-comp.\ column is the microscopic derivation from Eqs.~\eqref{eq:rde-three-comp} through the field-dependent rates $\alpha(\phi)$ and $\gamma(\phi)$; the 2 comp.\ column is its fast-reactivation limit $\nu\to\infty$. AMB~\cite{Wittkowski_Cates:2014} and phase-field crystal (PFC)~\cite{Elder_Grant:2002} are phenomenological field theories recovered as special cases of $\AMBm$, with $\eta^*(\phi) = -\phi + \phi^3$.%
}
\label{tab:params}
\end{table}

The local term $\eta^*(\phi)$ is the mass-redistribution potential evaluated on the reactive nullcline.
The gradient corrections arise because the steady-state profile lies on the plane ${\eta = \eta_\mathrm{stat}}$ rather than on the nullcline, and provide the corresponding deviation as an expansion in derivatives of $\phi$.
The interfacial coefficient ${\kappa(\phi) = D_{\mathrm{m}}/\gamma(\phi) - D_{\mathrm{c}}\alpha'(\phi)/\nu^2}$ comprises two contributions of distinct microscopic origin (End Matter).
The first, ${D_{\mathrm{m}}/\gamma(\phi) > 0}$, arises from the adiabatic elimination of the active cytosolic species and reduces in the fast-reactivation limit ${\nu \to \infty}$ to the Cahn--Hilliard stiffness coefficient of the corresponding two-component system~\cite{Weyer_Frey:2026, Zhou2026}.
The second, ${-D_{\mathrm{c}}\alpha'(\phi)/\nu^2 < 0}$, arises from the elimination of the inactive cytosolic species, whose gradient at the interface opposes that of the active component [Fig.~\ref{fig:model}(e)] and reduces the stabilising cytosolic flux that counteracts diffusive broadening of the membrane interface.

We specialise to cooperative recruitment, with $\gamma(\phi)$ growing with density, and to linear detachment, ${\alpha(\phi) = \alpha_0\phi}$.
With this choice ${\alpha''(\phi) = 0}$, so $\lambda$ vanishes and the biharmonic term $\nabla^4\chi(\phi)$ reduces to $K\nabla^4\phi$ with ${K = D_c \alpha_0 /\nu^2> 0}$. The nonlinear-detachment generalisation, in which a $\lambda(\phi)(\nabla\phi)^2$ contribution survives and is structurally identical to the active term of AMB, is deferred to future work.
The cooperativity of $\gamma$ makes the positive contribution to $\kappa(\phi)$ decrease with density, while the negative contribution remains constant; consequently $\kappa(\phi)$ changes sign at sufficiently high density. Above this density the steady-state profile is unstable to short-wavelength modulations, and the biharmonic term selects a finite wavelength.
In this specialisation, $\AMBm$ contains both the phase-field-crystal (PFC) model~\cite{Elder_Grant:2002} (equivalently, conserved Swift--Hohenberg), exactly obtained in the static-membrane limit ${D_m \rightarrow 0}$, and Model B (${\nu \rightarrow \infty}$) \cite{Hohenberg_Halperin:1977} as special cases (Table~\ref{tab:params}); $\AMBm$ continuously interpolates between these regimes, and we expect a mixed phenomenology spanning macrophase separation and finite-wavelength patterns.

\paragraph{Coexistence conditions.}
At steady state, the chemical potential is uniform, $\eta = \eta_\mathrm{stat}$ everywhere, as inherited from the mass conservation law shared by all McRD systems, irrespective of component number.
A second condition follows from the steady-state form of Eq.~\eqref{eq:AMBminus} for a planar interface, in the linear-detachment specialization,
\begin{equation}
  \eta_\mathrm{stat} = \eta^*(\phi)
  - \kappa(\phi)\partial^2_x\phi
  + K\partial^4_x\phi
  \label{eq:AMBminus-steadystate}
\end{equation}
on multiplication by $\partial_x\phi/\kappa(\phi)$ and integration across an interface connecting the plateaus $\phi_\pm$.

In the two-component limit (${\nu\to \infty}$), both active corrections vanish (${\lambda(\phi) \to 0}$ and ${\chi(\phi) \to 0}$, Table~\ref{tab:params}), and ${\kappa(\phi) > 0}$.
The integration yields an exact Maxwell-like construction:
\begin{align}
\label{eq:Maxwell-construction-del}
    \int_{\phi_-}^{\phi_+} \! d\phi \,
    \frac{\eta_\mathrm{stat}-\eta^*(\phi)}{\kappa(\phi)} = 0.
\end{align}
Since ${\kappa(\phi) > 0}$ throughout the interval, we may introduce a pseudo-density $\psi$ via ${d\psi = d\phi/\kappa(\phi)}$, with plateau values ${\psi_\pm \equiv \psi(\phi_\pm)}$, and a pseudo-potential $f_\psi$ via ${df_\psi = \eta^*(\phi)\,d\phi/\kappa(\phi)}$.
Equation~\eqref{eq:Maxwell-construction-del} integrates to ${\eta_\mathrm{stat}(\psi_+ - \psi_-) = f_\psi(\psi_+) - f_\psi(\psi_-)}$, equivalent to equality of a pseudo-pressures ${P = \eta_\mathrm{stat}\psi - f_\psi}$ on the two sides of the interface.
This construction has a direct physical interpretation: substituting the microscopic expressions for $\eta^*(\phi)$ and $\kappa(\phi)$ from Table~\ref{tab:params}, it recovers the reactive turnover balance of the underlying two-component reaction-diffusion system (End Matter).

At finite $\nu$, the interfacial coefficient $\kappa(\phi)$ can become negative, and a higher-order gradient term $\nabla^4 \chi(\phi)$ is required. The interplay between these two terms causes the pseudo-pressure construction to fail~\cite{Burekovic2026}, and the equal-pressure condition ceases to be an equation of state, becoming instead profile-dependent.
The coexistence conditions of $\AMBm$ reflect the original three-component model, where steady-pattern lie on to the plane of constant $\eta$ but the reaction turnover-balance is profile-dependent (see End Matter).
This places $\AMBm$ at a position within the landscape of scalar active field theories distinct from both $\AMBp$~\cite{Tjhung_Cates:2018}, where activity enters as a non-gradient current outside the $\nabla^2\eta$ structure breaking the chemical potential condition, and PFC~\cite{Elder_Grant:2002}, where constant $\kappa$ makes the Maxwell construction an exact equation of state.

\begin{figure}[!t]
\centering
\includegraphics[width=1\columnwidth]{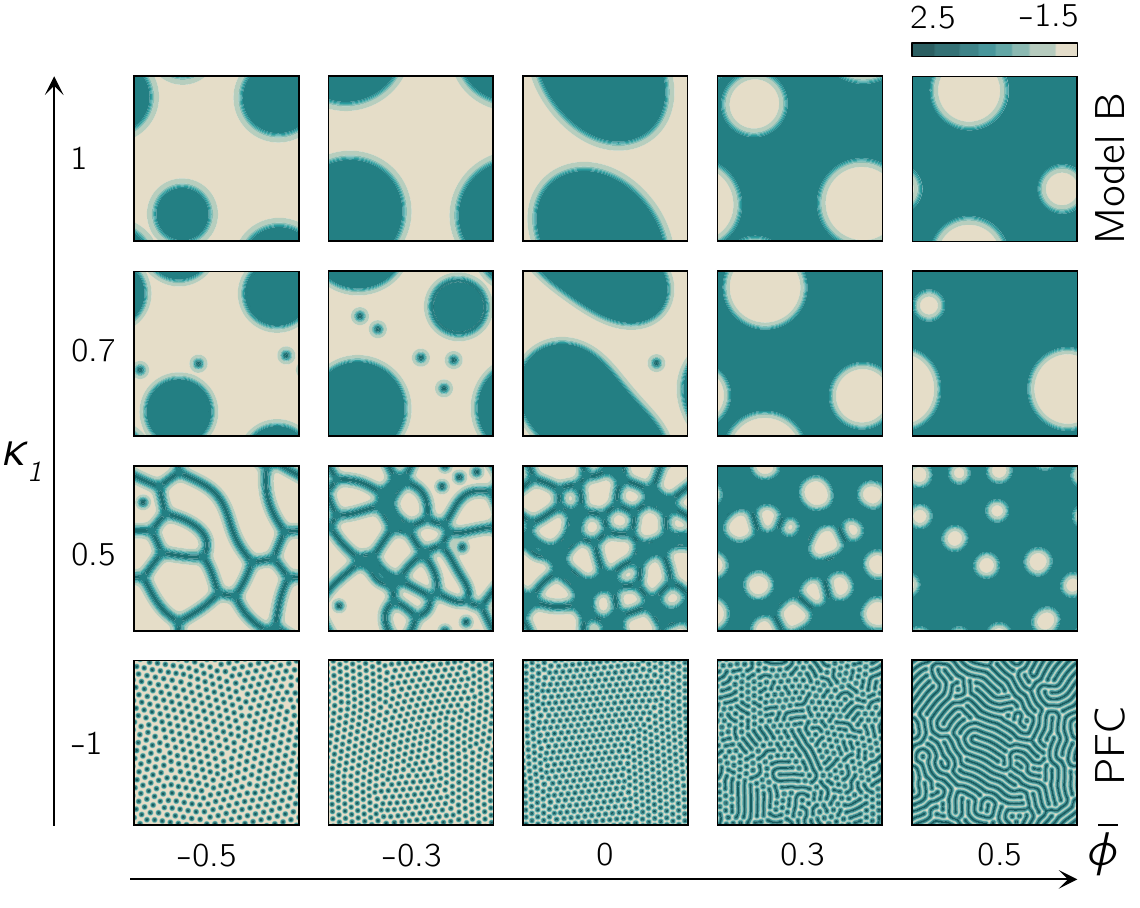}
\caption{
Phase diagram of phenomenological $\AMBm$, obtained from finite-element simulations~\cite{SM} in the $(\bar\phi, \kappa_1)$ plane. Increasing $\kappa_1$ at fixed mean density traverses a sequence of morphologies from the PFC limit (${\kappa_1 \lesssim 0}$, periodic patterns) to the Model B limit (${\kappa_1 \gtrsim 1}$, macrophase separation), passing through coexistence of large and small droplets, microphase-separated droplets, and stripes/foams in the intermediate regime. Colors encode local density $\phi$ (teal, dark gray; beige, light gray). Parameters: $\kappa(\phi) = \tanh(\kappa_1 - \phi)$, $M = 1$, $\chi = 0.055\phi$, $\eta^*(\phi) = -\phi + \phi^3$, $\lambda = 0$.
}
\label{fig:phasediagram}
\end{figure}

\paragraph{Simulation results.}
To test whether $\AMBm$ captures the finite-wavelength patterns of the three-component model, we performed two-dimensional finite-element simulations~\cite{COMSOL63,Toffenetti_etal:2026} of three systems: a phenomenological $\AMBm$ with ${\kappa(\phi)=\tanh(\kappa_1-\phi)}$, where $\kappa_1$ interpolates continuously between the PFC limit ($\kappa_1 \rightarrow -\infty$) and the Model~B limit ($\kappa_1 \rightarrow \infty$) [Fig.~\ref{fig:phasediagram}]; the full reaction--diffusion model Eq.~\eqref{eq:rde-three-comp} [Supplementary Fig.~1]; and the derived $\AMBm$ with microscopic parameters from Table~\ref{tab:params} [Supplementary Fig.~2]. The phase diagrams of the full three-component model and of the derived $\AMBm$ in the $(\bar\phi,\nu)$ plane agree quantitatively, and both reproduce qualitatively the sequence of morphologies found for the phenomenological $\AMBm$ in the $(\bar\phi,\kappa_1)$ plane [Fig.~\ref{fig:phasediagram}]: for large positive $\kappa_1$, the dynamics exhibits Cahn--Hilliard/Model-B coarsening; for large negative $\kappa_1$, the model reduces to the PFC, and the dynamics exhibit finite-wavelength patterns such as stripes and dots. In the crossover regime, where $\kappa(\phi)$ changes sign within the density range attained in the patterns, the phenomenological $\AMBm$ displays patterns absent from either limiting theory, including steady coexistence of small and large droplets at ${\kappa_1=0.7}$~\cite{Rossetto_Ernst_Zwicker:2025,Thewes_et_al:2025} [Supplementary Video~3] and foam-like states at ${\kappa_1=0.5}$. At high mean density, these foams coarsen by wetting [Supplementary Video~2], whereas at lower mean density they remain dynamically active and fail to settle into a stationary state [Supplementary Video~1], highlighting the nonvariational character of $\AMBm$. Foam-like structures also arise in the three-component and derived $\AMBm$ models, but there they always reach a stationary state, as in the Min system~\cite{Weyer_Frey:2026}.

\begin{figure}[t]
  \centering
\includegraphics[width=\columnwidth]{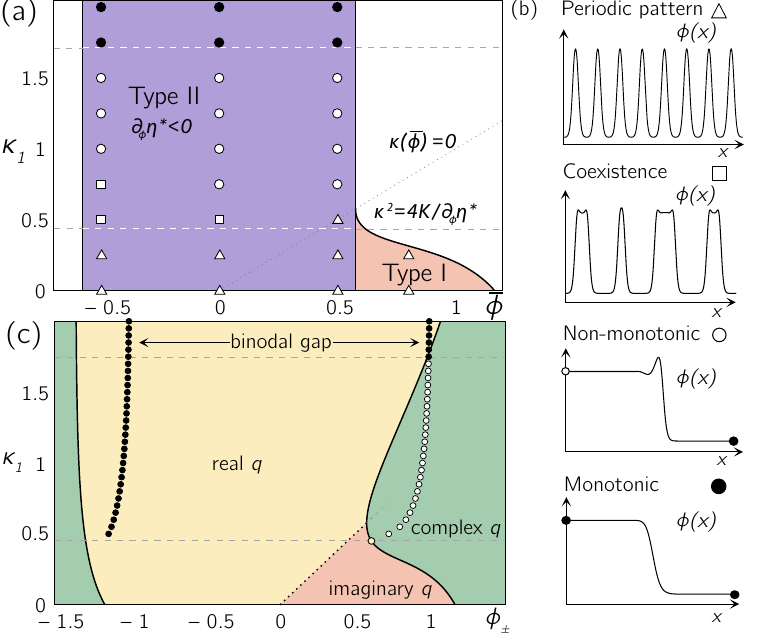}
\caption{%
(a) Linear stability of the homogeneous state in the $(\bar\phi,\kappa_1)$ plane (parameters as in Fig.~\ref{fig:phasediagram}): Type~II (lavender, dark gray) and Type~I (coral, medium gray) instabilities; stable elsewhere (white). Symbols in 1D simulations: triangles (periodic), squares (multi-wavelength), filled/open circles (monotonic/non-monotonic interfaces).
(b) Representative steady-state profiles, one per symbol class.
(c) Spatial-eigenvalue analysis of plateau solutions $\phi_\pm$, classified by the decay rate $q$: real (cream, light gray, monotonic tail), complex (mint, dark gray, non-monotonic tail), imaginary (coral, medium gray, plateau unstable). Solid: ${\kappa^2=4K\,\partial_\phi\eta^*}$; dotted: ${\kappa(\phi)=0}$. Circles: numerically extracted $\phi_\pm(\kappa_1)$, filled (open) inside the real-$q$ (complex-$q$) region; yellow filled circle: $\phi_+$ obtained by extrapolation.  Dashed lines: $\kappa_1$ values at which $\phi_+$ enters/exits complex $q$; arrow: binodal gap.
}
\label{fig:analytical-phase-diagram}
\end{figure}

\paragraph{One-dimensional analysis.}
To make analytical progress, we study $\AMBm$ in one dimension, where the elementary objects underlying the 2D phenomenology --- interfaces, periodic structures, and plateaus ---  remain accessible.
The 1D and 2D phase diagrams largely correspond [Figs.~\ref{fig:phasediagram} and~\ref{fig:analytical-phase-diagram}(a)], justifying the 1D reduction.
Linear stability analysis of the homogeneous state,
${\phi(x) = \bar\phi + \delta\phi\,e^{\sigma t + ikx}}$, yields
the dispersion relation ${\sigma(k) = -M k^2 [
  \partial_\phi\eta^*(\bar\phi)
  - \kappa(\bar\phi)\, k^2
  + K\, k^4]}$.
We find three regimes, in which the homogeneous state is stable, long-wavelength unstable (Type~II), or finite-wavelength unstable (Type~I)~\cite{Cross_Hohenberg:1993} [Fig.~\ref{fig:analytical-phase-diagram}(a)].
While only patterns with a finite wavelength form in the region of Type I instability, finite-wavelength patterns are observed in a much wider parameter regime.
In the simulations we find four distinct steady-state profiles [Fig.~\ref{fig:analytical-phase-diagram}(b)]:
monotonic interfaces between two plateaus, non-monotonic interfaces with spatial oscillations decaying onto the high-density plateau, periodic patterns of finite intrinsic wavelength, and coexistence of patterns with distinct wavelengths.
All four also appear in the three-component model [Supplementary Fig.~1].
In addition, $\AMBm$ admits localized states in 1D, which we expect to underlie the foam morphologies of [Fig.~\ref{fig:phasediagram}]; their systematic characterisation calls for continuation and bifurcation methods~\cite{Thiele2013}.

To obtain a criterion for the threshold between coarsening and interrupted coarsening, we linearise Eq.~\eqref{eq:AMBminus-steadystate} about the plateaus ${\phi=\phi_\pm}$ with the ansatz $\phi=\phi_\pm+\delta\phi\,e^{qx}$. The resulting quadratic in $q^2$ separates three regimes (End Matter), shown in Fig.~\ref{fig:analytical-phase-diagram}(c): real $q$ (monotonic decay), complex $q$ (non-monotonic decay), and imaginary $q$ (finite-wavelength replacement of the plateau).

Since no equation of state fixes $\phi_\pm$ in $\AMBm$, we extract them numerically [Fig.~\ref{fig:analytical-phase-diagram}(c)]. As $\kappa_1$ decreases, $\phi_+$ crosses from real to complex to imaginary $q$ while $\phi_-$ stays real; these crossings coincide with the morphology transitions of the 1D phase diagram [Fig.~\ref{fig:analytical-phase-diagram}(a)] and explain the profile types of (b). Multi-wavelength coexistence remains beyond this analysis.

The asymmetry of the two interface tails --- carried by
the field-dependent $\kappa(\phi)$ --- distinguishes $\AMBm$
from PFC and from other fourth-order gradient scalar
active-matter theories~\cite{Burekovic2026}. The $\kappa_1$
at which $\phi_+$ enters the imaginary-$q$ region coincides
with the Type-I instability region of
Fig.~\ref{fig:analytical-phase-diagram}(a), providing a
criterion for the onset of interrupted coarsening.

\paragraph{Discussion \& Outlook.}
$\AMBm$ is the minimal active scalar field theory showing microphase separation that admits a chemical
potential but no equation of state. It arises as the single-field projection
of a mass-conserving reaction--diffusion (McRD) system.
The mapping provides an explicit microscopic origin for the active contribution
$\lambda(\partial_x\phi)^2$ and the density-dependent interfacial coefficient
$\kappa(\phi)$ --- which can become negative at high density. The same structural absence of an
equation of state characterises multi-component McRD, where reaction--turnover
balance is profile-dependent while the mass-redistribution potential is
constant at steady state. This shared structure produces a shared
phenomenology: macrophase separation, droplet coexistence, and
wavelength-selected microphases (stripes, dots, foam) closely related to
patterns of the Min protein system~\cite{Frey_Weyer:2026}.

The mapping also raises two open questions. The profile-dependent $\eta_{\mathrm{stat}}$ controls pattern selection in 1D and is expected to do so in higher dimensions and for multi-wavelength coexistence;  whether projection of the well-characterised McRD noise yields a controlled noise prescription for $\AMBm$ is the second, an issue so far treated phenomenologically in scalar active field theories. More broadly, our results take a step toward unifying the languages of reaction--diffusion and active matter.

\begin{acknowledgments} We thank Daniel Zhou for stimulating and inspiring discussions.
This work was funded by the Deutsche Forschungsgemeinschaft (DFG, German Research Foundation) through the Excellence Cluster ORIGINS under Germany's Excellence Strategy (EXC-2094~--~390783311), the Excellence Cluster BioSystem under Germany's Excellence Strategy (EXC3092/1-533751719), the European Union (ERC, CellGeom, project number 101097810), and the Chan-Zuckerberg Initiative (CZI).
\end{acknowledgments}

\bibliographystyle{apsrev4-2}
\bibliography{literature}
\appendix

\section{End Matter}

\subsection{Reduction of the three-component model to $\AMBm$}

We give the algebraic steps underlying the reduction of
Eq.~\eqref{eq:rde-three-comp} to Eq.~\eqref{eq:AMBminus} in the
regime $\nu^{-1}, d_{\mathrm m} \ll 1$.
Mass conservation gives the continuity equation
\begin{equation}
  \partial_t \phi
  = D_{\mathrm c}\nabla^2 \eta ,
  \label{eq:phi-conservation-EM}
\end{equation}
so the task is to express $\eta$ as a functional of $\phi$.
We use $(\phi, \eta, c_{\mathrm i})$ as variables; the
definitions of $\phi$ and $\eta$ give
\begin{equation}
  m
  = \frac{\phi - \eta}{1 - d_{\mathrm m}},
  \qquad
  c_{\mathrm a}
  = \frac{\eta - d_{\mathrm m}\phi}{1 - d_{\mathrm m}}
    - c_{\mathrm i}.
  \label{eq:variables-transform-EM}
\end{equation}
Taking the time derivative of $\eta$ and using
Eq.~\eqref{eq:rde-three-comp} gives the exact balance
\begin{equation}
\begin{aligned}
  \partial_t \eta
  &= D_{\mathrm c}(1 + d_{\mathrm m})\nabla^2\eta
    - D_{\mathrm m}\nabla^2\phi \\
  &\quad
    - (1 - d_{\mathrm m})\bigl[\gamma(m)\,c_{\mathrm a} - \alpha(m)\bigr],
\end{aligned}
  \label{eq:eta-dynamics-EM}
\end{equation}
in which $m$ and $c_{\mathrm a}$ are understood through
Eq.~\eqref{eq:variables-transform-EM}.

\paragraph{Quasi-steady-state elimination of $\eta$.}
Near a stationary profile, $\eta$ is almost spatially uniform, while $\phi$ varies on the slow redistribution timescale.
We therefore set $\partial_t \eta \simeq 0$ and neglect the $\nabla^2 \eta$ term in Eq.~\eqref{eq:eta-dynamics-EM} ~\cite{Weyer_Frey:2023}.
Solving the resulting algebraic relation gives
\begin{equation}
  \eta
  = d_{\mathrm m}\phi
    + (1 - d_{\mathrm m})
      \left[
        c_{\mathrm i}
        + \frac{\alpha(m)}{\gamma(m)}
      \right]
    - \frac{D_{\mathrm m}}{\gamma(m)}\nabla^2\phi.
  \label{eq:eta-expanded}
\end{equation}
This relation is still implicit, since
$m = (\phi - \eta)/(1 - d_{\mathrm m})$ enters $\alpha(m)$ and
$\gamma(m)$.

\paragraph{Quasi-steady-state elimination of $c_{\mathrm i}$.}
The inactive pool relaxes on timescale ${\tau = 1/\nu}$, short
compared with the redistribution time of the conserved field.
Setting ${\partial_t c_{\mathrm i} = 0}$ in
Eq.~\eqref{eq:inactive-cytosol} gives the screened-Poisson
balance ${
  (\nu - D_{\mathrm c}\nabla^2) \, c_{\mathrm i}
  = \alpha(m)}$,
or equivalently, with ${\ell = \sqrt{D_{\mathrm c}\tau}}$,
\begin{equation}
  c_{\mathrm i}
  = \nu^{-1}
    (1 - \ell^2 \nabla^2)^{-1}\alpha(m).
  \label{eq:ci-resolvent-EM}
\end{equation}
For wavelengths large compared with the screening length,
$q\ell \ll 1$, the Neumann expansion yields
\begin{equation}
  c_{\mathrm i}
  = \nu^{-1}
    \left( 1 + \ell^2\nabla^2 + \ell^4\nabla^4 \right)
    \alpha(m)
    + \mathcal{O}(\nabla^6).
  \label{eq:ci-expanded}
\end{equation}
The fourth-order term is retained because it provides the leading
regularisation once the effective interfacial coefficient
becomes negative.

\paragraph{Closure on $\phi$.}
To close the dynamics on $\phi$ alone, we approximate the
membrane density in the local rates by the conserved density,
$m \to \phi$. Exactly, ${m = \phi - c_{\mathrm a} - c_{\mathrm i}}$,
so the closure is controlled where the cytosolic pools are small
compared with the membrane-bound density.
The inactive contribution is small in the fast-reactivation limit:
from Eq.~\eqref{eq:ci-expanded},
$c_{\mathrm i} = \mathcal{O}(\nu^{-1}\alpha)$, hence
$c_{\mathrm i}/m = \mathcal{O}(\nu^{-1})$ for detachment rates
that grow at most linearly with $m$ on the dense branch. The
approximation breaks down in the slow-reactivation regime
$\nu \to 0$, where the inactive pool can carry a finite fraction
of the total mass.
For the active component, stationarity gives
$\eta_{\mathrm{stat}} = c_{\mathrm a} + d_{\mathrm m} m
+ \mathcal{O}(\nu^{-1})$. Evaluated at the two plateaus,
this implies ${
  c_{\mathrm a, +} - c_{\mathrm a, -}
  = -\, d_{\mathrm m}(m_+ - m_-)
    + \mathcal{O}(\nu^{-1})}$,
so variations of $c_{\mathrm a}$ across the coexistence profile
are small compared with variations of $m$ when
$d_{\mathrm m} \ll 1$. For the rates used here,
$\alpha(m) \propto m$ and $\gamma(m) \propto 1 + m^2$, the dense
plateau scales as
$m_+ = \mathcal{O}(d_{\mathrm m}^{-1/2})$,
$c_{\mathrm a, +} = \mathcal{O}(d_{\mathrm m}^{1/2})$,
$c_{\mathrm i, +} = \mathcal{O}(\nu^{-1} d_{\mathrm m}^{-1/2})$,
and hence
$c_{\mathrm a, +}/m_+ = \mathcal{O}(d_{\mathrm m})$,
$c_{\mathrm i, +}/m_+ = \mathcal{O}(\nu^{-1})$. The closure is
therefore controlled on the dense side, where the negative
interfacial coefficient and the finite-wavelength instability
originate.
On the dilute side it is a leading-order scalar closure, which
remains accurate across the
parameter regimes studied here; Fig.~\ref{fig:model}(e)--(f)
shows a representative profile where $m\gg c_{\rm{i}}+c_{\rm{a}}$. 

\paragraph{Identification of the effective coefficients.}
Substituting Eq.~\eqref{eq:ci-expanded} into
Eq.~\eqref{eq:eta-expanded} and applying the closure
$m \to \phi$ gives $\eta$ as a gradient expansion in the conserved
field. Expanding consistently in $d_{\mathrm m}$ and retaining
terms to fourth order in gradients,
\begin{equation}
  \eta[\phi]
  = \eta^*(\phi)
    - \kappa(\phi)\nabla^2 \phi
    + \lambda(\phi)(\nabla\phi)^2
    + \nabla^4 \chi(\phi),
  \label{eq:eta-functional-EM}
\end{equation}
with the coefficient functions $\eta^*, \kappa, \lambda, \chi$
listed in Table~\ref{tab:params}. The interfacial coefficient
takes the form
\begin{equation}
  \kappa(\phi)
  = \frac{D_{\mathrm m}}{\gamma(\phi)}
    - \frac{D_{\mathrm c}\,\alpha'(\phi)}{\nu^2},
  \label{eq:EM-kappa}
\end{equation}
collecting one gradient contribution from each adiabatic
elimination. The positive piece $D_{\mathrm m}/\gamma(\phi)$ arises from the quasi steady-state slaving of $c_{\mathrm a}$ from Eq.~\eqref{eq:membrane-bound}, $c_{\mathrm a} = \alpha(m)/\gamma(m) - [D_{\mathrm m}/\gamma(m)]\nabla^2 m$, and encodes the deviation of $c_{\mathrm a}$ from its nullcline value $\alpha/\gamma$ after the closure $m\to\phi$. The negative piece $-D_{\mathrm c}\alpha'(\phi)/\nu^2$ comes from the elimination of $c_{\mathrm i}$.
Together with
Eq.~\eqref{eq:phi-conservation-EM}, Eq.~\eqref{eq:eta-functional-EM}
yields Eq.~\eqref{eq:AMBminus}.

\paragraph{Linear-$\alpha$ specialisation.}
For linear detachment $\alpha(\phi) = \alpha_0 \phi$, one has
$\alpha''(\phi) = 0$, so $\lambda(\phi) = \tau \ell^2 \alpha''
\equiv 0$, and $\nabla^4 \chi(\phi) = \nabla^4
[D^2_{\rm{c}}\alpha(\phi)/\nu^2]$ reduces to $K\nabla^4 \phi$ with
$K = D^2_{\rm{c}} \alpha_0/ \nu^2> 0$. Eq.~\eqref{eq:AMBminus} then
takes the canonical form used in the simulations of the main text.

\subsection{Steady-state structure of $\AMBm$}

At steady state $\eta = \eta_\mathrm{stat}$, the
profile $\phi(x)$ for ${\lambda=0}$ (linear detachment rate)  obeys the fourth-order ODE
\begin{equation}
  \eta^*(\phi) - \kappa(\phi)\,\phi'' + K\phi'''' = \eta_\mathrm{stat}\,.
  \label{eq:SS-ODE}
\end{equation}

\paragraph{Local linearisation: three regimes.}
The fixed points of Eq.~\eqref{eq:SS-ODE} satisfy
$\eta^*(\phi) = \eta_\mathrm{stat}$; for $\eta_\mathrm{stat}$ in
the relevant range there are three real roots, of which we
linearise around the outer two, $\phi_\pm$, identifying
them with the coexistence densities of the heteroclinic profile.
Linearising Eq.~\eqref{eq:SS-ODE} about a plateau density
$\phi_\pm$ with the ansatz
$\phi(x) = \phi_\pm + \delta\phi\,e^{q x}$ gives the
characteristic polynomial
\begin{equation}
  K q^4 - \kappa(\phi_\pm)\,q^2 + \partial_\phi\eta^*(\phi_\pm) = 0,
  \label{eq:SS-char}
\end{equation}
a quadratic in $q^2$. At the outer plateaus,
$\partial_\phi\eta^*(\phi_\pm) > 0$, so by Vieta's formulas
$q_1^2 + q_2^2 = \kappa(\phi_\pm)/K$ and
$q_1^2 q_2^2 = \partial_\phi\eta^*(\phi_\pm)/K$.
When the discriminant
$\kappa(\phi_\pm)^2 - 4K\partial_\phi\eta^*(\phi_\pm)$ is
positive, the two real roots in $q^2$ have the same sign, namely
the sign of $\kappa(\phi_\pm)$; when negative, the roots are
complex conjugate.
Three regimes follow.
\emph{(i) Real $q$} ($\kappa(\phi_\pm) > 0$,
$\kappa^2 > 4K\,\partial_\phi\eta^*$): both $q^2 > 0$, four real
roots, exponential decay to the plateau, recovering the standard monotonic
Cahn--Hilliard interface.
\emph{(ii) Complex $q$} ($\kappa^2 < 4K\,\partial_\phi\eta^*$):
$q^2$ complex conjugate, four roots $\pm a \pm ib$, decay to the
plateau with damped oscillations (non-monotonic interface). Because $\kappa(\phi)$ is
density-dependent, the two plateaus generically fall in
\emph{different} regimes, so one tail may oscillate while the
other does not --- a profile-internal asymmetry absent in PFC
and in fourth-order scalar gradient active-matter theories with
constant $\kappa$~\cite{Burekovic2026}.
\emph{(iii) Imaginary $q$} ($\kappa(\phi_\pm) < 0$,
$\kappa^2 > 4K\,\partial_\phi\eta^*$): both $q^2 < 0$, four
imaginary roots $q = \pm i\omega_{1,2}$ with
\begin{equation}
  2 K \omega_{1,2}^2
  = |\kappa(\phi_\pm)| \pm
    \sqrt{\kappa(\phi_\pm)^2 - 4K\,\partial_\phi\eta^*} \, .
  \label{eq:SS-omega}
\end{equation}
No spatially decaying mode connects to the plateau, so the
plateau cannot be approached as $x \to \pm\infty$.
The same conclusion is
reached from homogeneous LSA:
the plateau density itself lies inside the Type-I region of the
homogeneous LSA and is unstable to finite-wavelength
perturbations.

Note that the above discussion generalizes trivially to a non-zero $\lambda$-term, as this term contributes at order  $\mathcal{O}(\delta\phi)^2$ and therefore drops out of the linear-stability analysis.

\paragraph{Coexistence in PFC and two-component McRD.}
We consider a planar interface, so that $\phi$
depends only on the normal coordinate $x$ and Eq.~(\ref{eq:SS-ODE}) is an effective 1D ODE; this is the standard setting for Maxwell-like constructions, with curvature corrections being subleading in the sharp-interface limit.
For \emph{constant} $\kappa$ (PFC, $K > 0$), multiplying
Eq.~\eqref{eq:SS-ODE} by $\phi'$ makes every term a perfect
derivative. Integrating across the heteroclinic with
$\phi^{(n)} \to 0$ at both plateaus, the gradient terms drop
out, leaving the standard Maxwell construction
$\int_{\phi_-}^{\phi_+}d\phi\,[\eta^*-\eta_\mathrm{stat}] = 0$.
The PFC dynamics is gradient flow on the free energy
$\mathcal{F}[\phi] = \int dx\,\{f + \tfrac{1}{2}\kappa(\phi')^2
+ \tfrac{1}{2}K(\phi'')^2\}$ with $df/d\phi = \eta^*$, and the
bulk pressure $P = \phi\eta^* - f$ is a state function.

For $\phi$-dependent ${\kappa(\phi)>0}$ (two-component McRD,
${K=0}$), the multiplier $\phi'$ fails since $\kappa$ depends
on $\phi$. Multiplying Eq.~\eqref{eq:SS-ODE} by
$\phi'/\kappa(\phi)$ instead converts the square-gradient term
into a perfect derivative regardless of the $\phi$-dependence
of $\kappa$. Integrating across the interface gives
\begin{equation}
  \int_{\phi_-}^{\phi_+} \! d\phi \,
  \frac{\eta_\mathrm{stat}-\eta^*(\phi)}{\kappa(\phi)} = 0,
  \label{eq:pseudo-Maxwell}
\end{equation}
the phase-plane balance-of-areas condition of
Ref.~\cite{Brauns_Frey:2020}. The plateau densities $\phi_\pm$
are fixed by the endpoint values alone, independent of the
interface profile.

\paragraph{Coexistence in $\AMBm$.}
For \emph{sign-changing} $\kappa(\phi)$ ($\AMBm$, ${K > 0}$), the
multiplier $\phi'/\kappa(\phi)$ fails because $\kappa$ vanishes
inside the integration range. The natural extension is the
multiplier $\phi'/[\kappa(\phi)+\kappa_0]$, where $\kappa_0$ is
any constant chosen so that $\kappa(\phi)+\kappa_0 > 0$ across
the interface; such a $\kappa_0$ exists as the integration
spans a finite range of $\phi$. Splitting
$\kappa/[\kappa+\kappa_0] = 1 - \kappa_0/[\kappa+\kappa_0]$, the
unit term gives a perfect derivative that vanishes at the
plateaus, and the residual $\kappa_0/[\kappa+\kappa_0]$ term and
the biharmonic term combine to give the quasi-Maxwell
construction:
\begin{equation}
  \int_{\phi_-}^{\phi_+} \! d\phi \,
  \frac{\eta_\mathrm{stat}-\eta^*(\phi)}{\kappa(\phi)+\kappa_0} = \int_{-\infty}^{\infty} dx \,\phi'\,\frac{\kappa_0\phi''+K \phi''''}{\kappa(\phi)+\kappa_0}
  \label{eq:Maxwell-AMB-}
\end{equation}
$\kappa_0$ is an auxiliary shift, not a physical parameter:
Eq.~\eqref{eq:Maxwell-AMB-} holds as an identity for any
admissible $\kappa_0$, with both sides separately depending on
the choice. Unlike the PFC and two-component cases, the
right-hand side does not vanish; the osmotic- and
pseudo-pressure constructions both fail, and the coexistence
densities depend on the full interface profile.

\paragraph{Reaction turnover balance.}
The same condition expressed in $m$-coordinates,  before the
mapping to the scalar theory, reads as the reaction turnover balance: at
steady state, $c_{\mathrm{a}} = \eta_\mathrm{stat} - c_{\mathrm{i}} - d_{\mathrm{m}} m$;
substituting into the stationary membrane
equation~\eqref{eq:membrane-bound} (with a linear detachment rate $\alpha(m)=\alpha_0 m$), multiplying by
$\partial_x m$, and integrating across the interface yields
\begin{equation}
\begin{split}
  \int_{m_-}^{m_+}\!dm\,
  \bigl[\gamma(m)(\eta_\mathrm{stat} - d_{\mathrm{m}} m) - \alpha_0\,m\bigr]
  \\
  = \int_{-\infty}^{+\infty}\!dx\;c_{\mathrm{i}}(x)\,\gamma(m)\,\partial_x m,
\end{split}
\label{eq:turnover-balance}
\end{equation}
the $m$-coordinate version of Eq.~\eqref{eq:Maxwell-AMB-}.
Equivalence with Eq.~\eqref{eq:Maxwell-AMB-} follows from
three identifications, consistent with Table~\ref{tab:params}:
\begin{equation}
\begin{aligned}
  \eta^*(\phi) &=  \alpha_0\,\phi/\gamma(\phi)+\alpha_0\,\phi/\nu+d_{\mathrm{m}} \phi, \\
  \kappa(\phi) + \kappa_0 &= D_{\mathrm{m}}/\gamma(\phi), \\
  c_{\mathrm{i}}(\phi) &=\phi \,\alpha_0/\nu+ \kappa_0\,\phi'' + K\,\phi'''' + \cdots,
\end{aligned}
\label{eq:turnover-equivalence}
\end{equation}
which fix $\kappa_0 = D_{\mathrm{c}}\alpha_0/\nu^2$. The first
relation defines $\eta^*$ at the nullcline; the second
identifies the integration measure $1/[\kappa+\kappa_0] =
\gamma(\phi)/D_{\mathrm{m}}$; the third recasts the gradient
expansion of $c_{\mathrm{i}}$, Eq.~\eqref{eq:ci-expanded}, as
the integrand on the right-hand side of
Eq.~\eqref{eq:Maxwell-AMB-}. Inserting these three identifications into Eq.~\ref{eq:Maxwell-AMB-} recovers
the turnover balance Eq.~\eqref{eq:turnover-balance} up to a change of variable
$m \rightarrow \phi$, consistent with our closure scheme.

In the two-component limit $\nu\to\infty$ the right-hand
side of Eq.~\eqref{eq:turnover-balance} vanishes, recovering
Ref.~\cite{Brauns_Frey:2020}; finite $\nu$ retains the explicit
profile dependence.

\end{document}